\documentclass[aps,prl,twocolumn]{revtex4}
\usepackage{xspace}
\usepackage{float}
\usepackage{amsmath}
\usepackage[normalem]{ulem}

\usepackage{amssymb,amssymb,psfrag,epsfig,graphicx}

\newcommand{\fig}[2]{\includegraphics[width=#1]{./figures/#2}}

\newlength{\bilderlength}

\newcommand{\rmd}{{\mathrm{d}}}

\begin{document}
\bibliographystyle{KAY}
%%%%%%%%%%%%%%%%%%%%%%%%%%%%%%%%%%%%%%%%%%%%%%%%%%%%%%%%%%%%%%%%%%%%%%%%%%%%
\title{\sffamily \bfseries \Large Numerical Calculation of the
Functional renormalization group fixed-point functions at the
depinning transition}

\author{Alberto Rosso} \affiliation{LPTMS; CNRS and Universit\'e
Paris-Sud, UMR 8626, ORSAY CEDEX 91405, France.}

\author{Pierre Le Doussal}\affiliation{ CNRS-Laboratoire de
Physique Th{\'e}orique de l'Ecole Normale Sup{\'e}rieure, 24 rue
Lhomond, 75231 Paris Cedex, France.}

\author{Kay J\"org Wiese}
\affiliation{ CNRS-Laboratoire de
Physique Th{\'e}orique de l'Ecole Normale Sup{\'e}rieure, 24 rue
Lhomond, 75231 Paris Cedex, France.}

\date{\small\today}
\begin{abstract}
We compute numerically the sequence of successive pinned
configurations of an elastic line pulled quasi-statically by a spring
in a random bond (RB) and random field (RF) potential.  Measuring the
fluctuations of the center of mass of the line allows to obtain the
functional renormalization group (FRG) functions at the depinning
transition. The universal form of the second cumulant $\Delta(u)$ is
found to have a linear cusp at the origin, to be identical for RB and
RF, different from the statics, and in good agreement with 2-loop
FRG. The cusp is due to avalanches, which we visualize. Avalanches also
produce  a cusp in the third cumulant, whose universal form is
obtained, as predicted by FRG.
\end{abstract}
\maketitle

Universality is often more difficult to characterize in random systems
than in their pure counterparts. Sample-to-sample fluctuations
complicate the analysis and the nature of the critical theory may be
different. One prominent example is the zero-temperature ($T=0$)
depinning transition from a pinned to a moving state, which occurs
when an interface is pulled through a random medium by an external
force $f$ beyond a threshold $f_c$. Its understanding is important for
magnets \cite{domain-walls-exp}, ferroelectrics \cite{ferroelectrics},
super-conductors \cite{zeldov,review_pinning_russes}, density waves
\cite{brazovskii03}, wetting \cite{rolley}, dislocation \cite{moretti}
and crack propagation \cite{ebouchaud}, and earthquake dynamics
\cite{fisher-phys-rep98}. At the transition the interface displacement
$u(x)$ is expected to scale as $u(x)-u(0) \sim x^\zeta$, where $x$ is
the $d$-dimensional internal coordinate and $\zeta$ the roughness
exponent. The analogy with critical phenomena, suggested by mean-field
theory \cite{fisher85}, was analyzed using the functional
renormalization group (FRG) to one loop
\cite{depinning2,narayan-fisher93}. 2-loop FRG studies resolved the
apparent contradiction that statics ($f=0$) and depinning ($f=f_c$)
cannot be distinguished at one loop
\cite{ChauveLeDoussalWiese2000a,LeDoussalWieseChauve2002}. In the
earlier works \cite{fisher85,narayan-fisher93}, the presence of a
diverging length scale was argued to lead to the universal behavior
observed at the transition. This correlation length was observed in
numerics for the steady-state dynamics above $f_c$, but only in
transients below $f_c$
\cite{middleton_deppining_below_fc,chen_marchetti,duemmer,kolton_below}.
The FRG study \cite{creep} of thermally activated motion below
threshold showed a more complex picture with additional length scales
involving both statics and depinning. This is in agreement with a
recent numerical analysis of the $T=0^+$ steady state in that regime;
\cite{kolton06} shows that (i) there are no geometric diverging length
scales at $f_c^-$ for this steady state and (ii) the roughness is
given by the equilibrium static exponent at small scales and by the
depinning exponent at large scales for all $0< f < f_c$. The physics
is thus more subtle than in standard critical phenomena.  The 2-loop
FRG is a good candidate to describe this physics as it contains a
mechanism for a crossover between statics and depinning directly in
the quasi-static limit $T,v \to 0$ (by the generation of an anomalous
term in the $\beta$-function at any $f>0$). It is thus important to
directly test the central ingredients and properties of the FRG
approach in the dynamics, making contact with observables beyond
critical exponents.

Recently a method to measure the fixed-point function of the FRG for
the statics of pinned manifolds was proposed \cite{pld}. Exact
numerical determination of ground states for interfaces in various
types of disorders \cite{alan} shows a remarkable agreement in the
statics between the measured renormalized pinning-force correlator,
$\Delta(u)$, and the 1- and 2-loop predictions from the FRG
\cite{DSFisher1986,ChauveLeDoussalWiese2000a,statics2loop}. This
method has been extended to the quasi-static depinning
\cite{pld_kay}. The aim of the present paper is to compute numerically
these fixed-point functions for depinning. Outstanding predictions of
the FRG which we test here are: The existence of a linear cusp for
$\Delta(u)$, a single universality class for both random-bond (RB) and
random-field (RF) disorder, the difference of $\Delta (u)$ between the
static and depinning fixed points, and a comparison with 1- and 2-loop
predictions. In addition we study the third cumulant, which also
exhibits a cusp. The cusps in these FRG fixed-point functions can
directly be related to ``avalanches'' or ``dynamical shocks''.

\newlength{\figsize}
\setlength{\figsize}{1\columnwidth}
\begin{figure}[t]\setlength{\unitlength}{1.4mm}
\fboxsep0mm \psfrag{x}[][]{\small $2 \sin(q/2)$}
\psfrag{y}[][]{\small $S(q)$}
\psfrag{depinning-slope-zetauxyz}[][]{\small depinning slope
$\zeta=1.26$} \psfrag{massuxyz}[][]{\small $m=0.01$}
\psfrag{0.07}[][]{\small $0.07$} \psfrag{0.10}[][]{\small $0.10$}
\psfrag{0.20}[][]{\small $0.20$}\psfrag{0.50}[][]{\small $0.50$}
\mbox{\fig{\figsize}{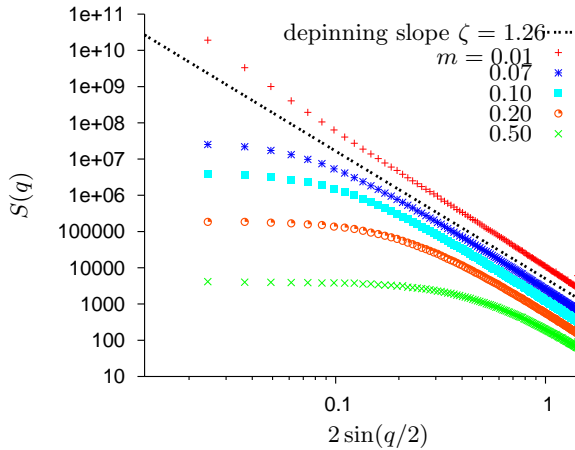}} \caption{Structure factor of the
line ($L=512$) for RF disorder (curves are shifted for clarity). The
crossover between the depinning slope and the plateau determines the
correlation length ($\propto m^{-1}$). For $m=0.01$ the correlation
length is bigger than the system size.} \label{structure}
\end{figure}

The main idea to study depinning, described in \cite{pld_kay}, is to
put the system in a quadratic potential and to move its center,
denoted $w$, monotonously and quasi-statically: The difference between
the center of mass of the manifold and $w$ will fluctuate, and its
second cumulant yields precisely the function $\Delta(u)$ defined and
computed in the FRG. In the continuum, the zero-temperature
Langevin dynamics is described by the equation of motion:
\begin{eqnarray}\label{eq:eqmo}
 \partial_t u(x,t) &=& {\cal F}_{w(t)}(x,u(x,t)) \\
 {\cal F}_{w}(x,u(x)) &=& m^2 (w - u(x)) + c \nabla^2 u(x) +
F(x,u(x)),\quad\nonumber
\end{eqnarray}
where ${\cal F}_{w}(x,u)$ is the total force acting on the manifold,
$c$ is the elastic constant and $m^2$ is the curvature of the
quadratic potential which acts as a mass for the field $u$. $F(x,u)$
is the random pinning force. For RF disorder, $F(x,u)$ is
short-ranged with correlations $\overline{F(x,0) F(x',u)}=
\Delta_0(u) \delta^{d}(x-x')$. For RB disorder this force is derived
from a short-ranged random potential $V(x,u)$, $F(x,u)= -\partial_u
V(x,u)$. % The motion can be described as follows.

Starting from an arbitrary initial condition $u_{\mathrm{init}}(x)$,
and a given $w=w_0$, the manifold moves to a locally stable state
$u_{w_0}(x)$, i.e.\ a zero-force state ${\cal
F}_{w_0}(x,u_{w_0}(x))=0$, which is stable to small deformations.
Increasing $w$,  $u(x)$ increases slightly (and smoothly if $F(u,x)$
is smooth in $u$), while the configuration remains stable. At some
$w=w_1$, the state becomes unstable and the manifold moves until it
is blocked again in a new locally stable state $u_{w_1}(x)$.  We are
interested in the center of mass (i.e.\ translationnally averaged)
displacement $u(w) = L^{-d} \int \rmd^d x\, u_w(x)$. The function
$u(w)$ exhibits jumps at a discrete set of values of $w$ and is in
general dependent on the initial condition. However, due to the
no-passing rule \cite{middleton}, we can prove that there exists a
$w^*>w_0$ such that the orbit $u_{w>w^*}(x)$ becomes independent of
the initial condition $u_{\mathrm{init}}(x)$, and $w_0$. A
stationary state is thus reached after a finite $w-w_0$, on which we
focus.

We  check these predictions numerically for a string ($d=1$).  To solve
 Eq.~(\ref{eq:eqmo}), we discretize the string along the $x$
direction, $x \rightarrow j=0,\ldots,L-1$, keeping $u_w(j)$ as a
continuous variable. A very efficient algorithm \cite{rosso1} finds
the exact location of the succession of locally stable states. For
RB disorder, we generate, for each $j$, a cubic spline $V(j,u(j))$
interpolating a large number ($10^2-10^3$) of uncorrelated normal
random points, of regular spacing $a$, zero derivative being imposed
at the first and the last points. Once $u(j)$ passes the last point,
a new spline is generated. For RF disorder, $F(j,u(j))$ is taken as
a linear interpolation of regularly spaced normal random points. The
discretization in $x$, in the limit $a \to 0$, preserves the
statistical tilt symmetry (STS) of the continuum model. (Only
very small corrections to $c$ are expected as $m a/\sqrt{c} \ll 1$).
The Fourier modes and center of mass of the discretized line are
defined as $u_q= \sum_{j=0}^{L-1} e^{i q j } u_w(j)$ and
$u(w)=u_0/L$.

\begin{figure}[t]\setlength{\unitlength}{1.4mm}
\fboxsep0mm \psfrag{x}[][]{\small $ \sqrt{m \Delta(0)}$}
\psfrag{y}[][]{\small $f_c(m)$} \psfrag{l1}[][]{\small
$f_c^{\text{RF}} \sim 0.7$ } \psfrag{l2}[][]{\small $f_c^{\text{RB}}
\sim 1.9$} \psfrag{RFxLy32zzz}[][]{\small RF $L=32$}
\psfrag{RBxLy32zzz}[][]{\small RB $L=32$}
 \psfrag{fit}[][]{\small fit} \psfrag{128}[][]{\small $128$}
 \mbox{\fig{\figsize}{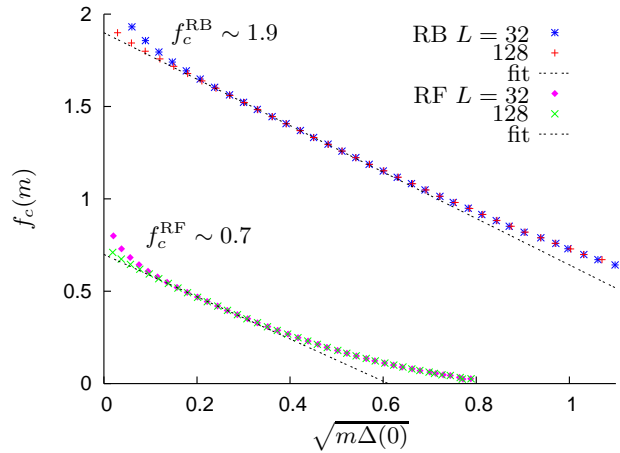}}
\caption{Finite-size study of $f_c(m)$. The extrapolation
$f_c=f_c(0)$ corresponds to the critical force for the infinite
system.} \label{force}
\end{figure}

\begin{figure}[b]\setlength{\unitlength}{1.4mm}
\fboxsep0mm   
\psfrag{random-field-disorder-num}[][]{\small RF $m=0.071$, $L=512$}
\psfrag{random-bond-disorder-num}[][]{\small RB $m=0.071$, $L=512$}
\psfrag{y}[][]{\small $Y (z)$}
\psfrag{x}[][]{\small $z$} 
\mbox{\fig{\figsize}{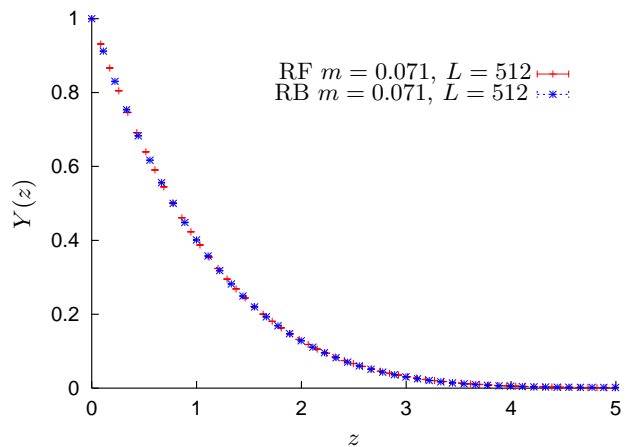}}
\caption{Universal scaling form $Y (z)$ for $\Delta (u)$ for RB and RF
disorder.} \label{f:Delta}
\end{figure}

We have observed that in the transient regime $w-u(w)$ increases on
average linearly with $w$ and reaches a plateau in the stationary
state. There, the line is depinning-like rough up
to a scale of order $1/m$ where the confinement due to the mass takes
over. This is apparent on the disorder-averaged structure factor
$S(q)=\overline{u_q u_{-q}}$ plotted in Fig.~\ref{structure} for
various masses: it exhibits a plateau at small $q$. In the steady
state the fluctuations of $w-u(w)$ are related to the FRG
functions. The first cumulants are \cite{pld_kay}:
\begin{eqnarray}\label{defDelta}
&& m^2\, \overline{[w - u(w)]} = f_c(m) \\
&& m^4\, \overline{[w - u(w)][w' - u(w')]}^c = L^{-d}
\Delta_m(w-w').\qquad\nonumber
\end{eqnarray}
Since the correlations of $w-u(w)$ decay over a finite scale in $w$,
the disorder averages in (\ref{defDelta}) can be determined as
translational averages over $w$. A prediction of the FRG is that in
the limit $m L \to \infty$ the quantities $f_c(m)$ and $\Delta_m(w)$
in (\ref{defDelta}) become $L$-independent. Here
Fig.~\ref{structure} shows that this holds for $L m \ge 5 \sqrt{c}$,
as one can check that several modes are confined and the correlation
length is smaller than the system size.  The FRG also predicts that:
\begin{eqnarray}
&& f_c(m)=f_c + c_1 m^{2-\zeta} \\
&& \Delta(u)= m^{\epsilon - 2 \zeta} \tilde \Delta(u m^\zeta)
\label{scaling}
\end{eqnarray}
where $\tilde \Delta(w)$ goes to a fixed point as $m \to 0$
($\epsilon=4-d$; here $d=1$).

\begin{figure}[t]\setlength{\unitlength}{1.4mm}
\fboxsep0mm \psfrag{x}[][]{\small $z$} \psfrag{y}[][]{\small
$Y(z)-Y_1(z)$} \psfrag{2loop-dynamique}[][]{\small $2$--loop dynamics}
\psfrag{2loop-statique}[][]{\small $2$--loop statics}
\psfrag{dynamique}[][]{\small $\frac{\rmd}{\rmd \epsilon
}Y^{\text{dyn.}}_{\epsilon}(z)$} \psfrag{statique}[][]{\small
$\frac{\rmd}{\rmd \epsilon }Y^{\text{stat.}}_{\epsilon}(z)$}
\psfrag{random-field-disorder-num}[][]{\small RF $m=0.071, L=512$}
\psfrag{random-bond-disorder-num}[][]{\small RB $m=0.071, L=512$}
\mbox{\fig{\figsize}{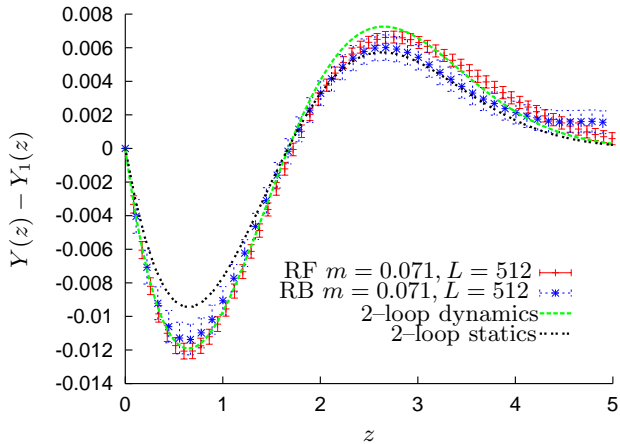}} \caption{The difference between the
normalized correlator $Y(z)$ and the 1-loop prediction
$Y_1(z)$. Averages over $10^7$ - $10^{8}$ samples were performed.}
\label{RF-RB}
\end{figure}

We have studied the behavior of the critical force $f_c(m)$ for the
two classes of disorder. From (\ref{scaling}) one has $\sqrt{\Delta(0)
m} \sim m^{2-\zeta}$, yielding a parameter-free linear scaling shown
in Fig.~\ref{force}. For large $m$ the scaling is non-linear, while
for smaller $m$ it is linear up to the scale where the correlation
length becomes of the order of $L$ ($m L\approx 5$). The critical
force of the infinite system is defined here in an unambiguous way, as
$f_{c}=f_{c} (m=0)$.  The resulting $c_1 < 0$; hence the average force
exerted on the manifold is smaller than $f_{c}$. One can see on
Fig.~\ref{force} that the slope of the two curves coincides. This is
consistent with the FRG which  predicts that it is a universal
amplitude, depending on microscopic details only through the
renormalized elastic coefficient $c$; here $c\approx 1$ for both
models of Fig.~\ref{force}. The study of this and related
amplitudes is deferred to a future publication. Here we focus  on
parameter-free fully universal functions.

We now turn to the determination of the fixed-point function. Since there are
two scales in $\Delta(u)$, we write
\begin{eqnarray}
 \Delta(u)= \Delta(0) Y(u/u_{\xi}),
\end{eqnarray}
where $Y(0)=1$ and one determines $u_\xi$ such that $\int_{0}^{\infty}
\rmd z\, Y(z)=1$. The function $Y(z)$ is universal and depends only on
the dimension of space. We have determined $Y(z)$ from our numerical
data both for RF and RB disorder. For small masses the two functions
coincide within statistical errors. This is visualized in
Fig.~\ref{f:Delta}. The prediction from the FRG is that $Y(z)=Y_1(z)+
\epsilon Y_2(z) + O(\epsilon^2)$ with $\epsilon=4-d$. The 1-loop
function is the same as for the statics and given by the solution of
$\gamma z=\sqrt{Y_{1}-1-\ln Y_{1}}$ and $\gamma = \int_{0}^{1} \rmd y
\, \sqrt{y-\ln y-1} \approx 0.5482228893$. Since the measured $Y(z)$
is numerically close to $Y_1(z)$, as in the statics, we plot in
Fig.~\ref{RF-RB} the difference $Y-Y_1$. The overall shape of the
difference is very similar to the one obtained for the RF statics in
$d=3,2,0$, which exhibits only a weak dependence on $d$. However the
overall amplitude is larger by a factor of about $1.25$, both in the
numerics and in the 2-loop FRG. We have plotted the function
$Y_2(z)=\frac{\rmd}{\rmd\epsilon} Y(z)|_{\epsilon=0}$ which, as for
the statics, is close to the numerical result.  We also observe a
cusp, i.e.\ $Y'(0)=-0.816 \pm 0.004$ for RF and $Y' (0) = -0.815 \pm
0.005$ for RB. FRG predicts $Y'(0)= -0.775304 - 0.0412061 \epsilon+O
(\epsilon^{2})$.

\begin{figure}[t]\setlength{\unitlength}{1.4mm}
\fboxsep0mm \psfrag{x}[][]{\small $x=\Delta(w)/\Delta(0)$}
\psfrag{y}[][]{\small $Q (x)$}
%\psfrag{1-loop}[][]{\small $1$--loop} \psfrag{l2}[][]{\small RF}
\psfrag{1loop-dynamique}[][]{\small $1$--loop prediction}
\psfrag{random-field-disorder-num}[][]{\small RF $m=0.071, L=512$}
\psfrag{random-bond-disorder-num}[][]{\small RB $m=0.071, L=512$}
%\psfrag{l3}[][]{\small RB}
\mbox{\fig{\figsize}{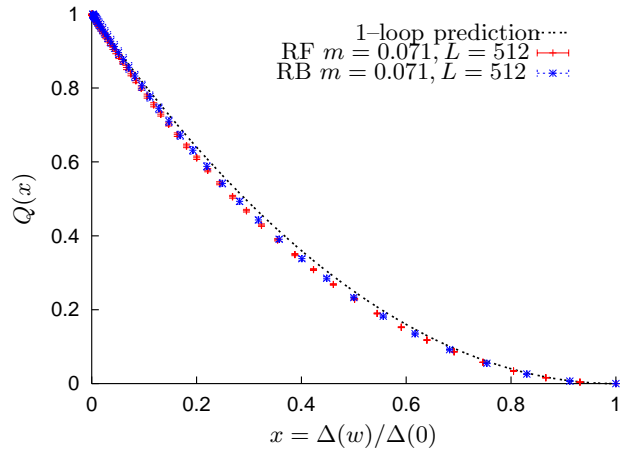}} \caption{Data collapse for the
universal function $Q (x)$ defined in (\ref{Q}), for RB and RF
disorder.
%Simulations has been performed for systems of size $L=512$ and mass
%$m=0.071$.
The line represents the $1$-loop prediction $Q(x)=(1-x)^2$.}
\label{3points}
\end{figure}

To investigate deeper the validity of FRG, we measure the third
cumulant \footnote{The corresponding quantity in the statics was
measured in \cite{alan}.}, defined as
\begin{equation}\label{third-cumul}
 m^{6} \overline{(w'- u(w')-(w-u(w)))^3}^c = L^{-2 d} S(w'-w)\ .
\end{equation}
The lowest order prediction \cite{pld} is $S(w)=\frac{12}{m^2}
\Delta'(w) (\Delta(0)-\Delta(w))$. To check the scaling in a
parameter-free way, we define
\begin{equation}\label{Q}
Q(\Delta(w)/\Delta(0)) := \int_0^w S(w') \rmd w'\Big/\int_0^\infty
S(w')\rmd w'.
\end{equation}
 $Q(x)$ is expected to be universal. Indeed we find, that RB and RF
give, within statistical errors, identical results, close to the
1-loop prediction, see Fig.~\ref{3points}.

\begin{figure}[t]\setlength{\unitlength}{1.4mm}
\fboxsep0mm \psfrag{y}[][]{\small $w{-}u{-}\overline{(w{-}u)}$}
\psfrag{x}[][]{\small $w{-}\overline{(w{-}u)}$}
\psfrag{randomfieldm03l8}[][]{\small  $m=0.3$, $L=8$}
\psfrag{randomfieldm04l8}[][]{\small  $m=0.4$, $L=8$}
\psfrag{randomfieldm05l8}[][]{\small  $m=0.5$, $L=8$}
\mbox{\fig{\figsize}{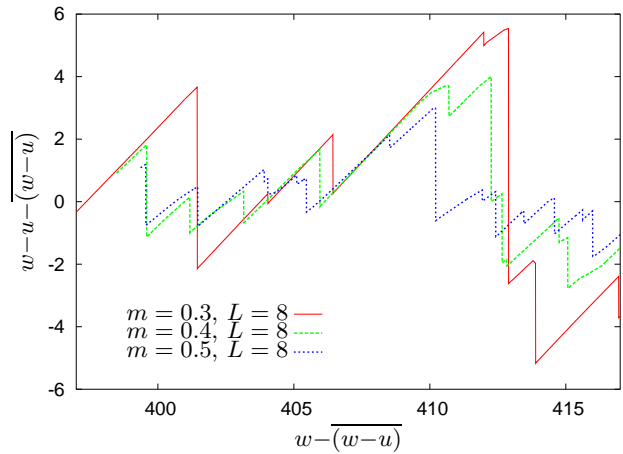}} \caption{Shocks (jumps) in the evolution
of the center of mass. Decreasing $m$, several small shocks merge into
few bigger ones.}  \label{f:shocks}
\end{figure}

It is instructive to visualize the shape of the function $u(w)$ in a
single environment as a function of the mass $m$, as shown in
Fig.~\ref{f:shocks}. The analogous function in the statics, i.e.\ the
position of the center of mass of the ground state, exhibits shocks.
In $d=0$ the evolution of these shocks as $m$ is decreased follows a
ballistic aggregation process described by the Burgers equation. The
``dynamical shocks'' or avalanches also follow an interesting dynamics
which remains to be studied in detail. The fact that they do not
accumulate, apparent on Fig.~\ref{f:shocks}, is consistent with the
linear cusp found in the second and third cumulants.

To conclude we have analyzed the dynamics of a manifold at the
depinning transition in a geometry which allows a precise and
unambiguous comparison to the predictions from the Functional RG. By
moving the quadratic well quasi-statically, we cleanly define the
avalanches at the threshold. The center-of-mass fluctuations become
universal and are described by the FRG fixed-point functions. The main
and non-trivial predictions of 2-loop FRG are confirmed, namely a
scale-invariant fixed-point function $\Delta(u)$ with a linear cusp
and a single universality class for RB and RF disorder. The precision
of the data allows for a quantitative comparison with the statics, in
agreement with FRG. A more detailed analysis of other universal
observables is deferred to a future publication
\cite{rosso_future}. We hope that this study opens the way to more
precise comparisons between theory and experiments.

We thank A.\ Fedorenko, W.~Krauth and A.\ Middleton for useful
discussions, and the KITP for hospitality. This work is supported by
ANR (05- BLAN-0099-01).

\end{document}